# The Low-Temperature Fate of the 0.7 Structure in a Point Contact: A Kondo-like Correlated State in an Open System

S. M. Cronenwett[1,2], H. J. Lynch[1], D. Goldhaber-Gordon[1,2],

L. P. Kouwenhoven[1,3], C. M. Marcus[1], K. Hirose[4], N. S. Wingreen[5], V. Umansky[6]

[1] Department of Physics, Harvard University, Cambridge, MA 02138

[2] Department of Physics, Stanford University, Stanford, CA 94305-4060

[3] Department of Applied Physics and ERATO, Delft University of Technology, PO Box 5046, 2600 GA Delft, Netherlands

[4] Fundamental Research Laboratories, NEC Corporation, 34 Miyukigaoka, Tsukuba, Ibaraki 305-8501, Japan

[5] NEC Research Institute, 4 Independence Way, Princeton, New Jersey 08540

[6] Braun Center for Submicron Research, Weizmann Institute of Science, Rehovot 76100, Israel

Besides the usual conductance plateaus at multiples of $2e^2/h$, quantum point contacts typically show an extra plateau at ~ $0.7(2e^2/h)$, believed to arise from electron-electron interactions that prohibit the two spin channels from being simultaneously occupied. We present evidence that the disappearance of the 0.7 structure at very low temperature signals the formation of a Kondo-like correlated spin state. Evidence includes a zero-bias conductance peak that splits in a parallel field, scaling of conductance to a modified Kondo form, and consistency between peak width and the Kondo temperature.

The quantization of conductance in units of $2e^2/h$ of a quantum point contact (QPC), a narrow constriction formed in a clean two-dimensional electron gas, has become a paradigm of mesoscopic physics [1,2]. This quantization indicates full transmission of the one-dimensional (1D) modes of the constriction, with the factor of two reflecting the spin degeneracy of the modes. Further support for this simple picture is the appearance of plateaus at odd multiples of $e^2/h$ in a large magnetic field—evidence of Zeeman splitting of the 1D modes [2,3]. Recently, the existence of an additional plateau around $0.7(2e^2/h)$, which becomes more prominent as the temperature is *increased*, has been investigated by many groups [4-9]. This feature, termed "0.7 structure," appears to evolve continuously out of the lowest spin-resolved plateau at $e^2/h$ as in-plane magnetic field, B, is lowered to zero [4]. We emphasize that 0.7 structure is observed more often than not, and is evident even in the earliest experiments on QPCs [1,3,10].

The first reported systematic investigation of the 0.7 structure in a QPC proposed that this feature arises from a spontaneous spin polarization, presumably due to electron-electron interactions [4]. Since then, several other experiments on QPCs [5-8] and clean quantum wires [8,9] have provided further evidence connecting the 0.7 structure at zero magnetic field with known spin polarization effects at higher magnetic field. Several theoretical models have found a breaking of spin degeneracy in QPCs at low electron density [11], though no microscopic model has yet shown the 0.7 structure emerging directly from electron-electron interactions.

The fact that the 0.7 structure becomes stronger at higher temperature suggests that the feature is not a ground-state property. Instead, the 0.7 structure appears as a crossover from perfect conductance at low temperature to a reduced conductance at higher temperature. In systems with a spin degree of freedom, such a crossover is the hallmark of the Kondo effect—the screening of a localized spin by the formation of singlet correlations with the Fermi sea at low temperature. The appearance of Kondo-like effects in a QPC suggests a lifted degeneracy in the QPC, presumably resulting from Coulomb energy, that gives rise to a dynamic unpaired spin [12], rather than a static magnetic moment.

In this Letter, we present experimental evidence that at low temperature the unpaired spin associated with the 0.7 structure forms a Kondo-like, correlated many-body state. We find a number of similarities between the present system and the Kondo effect seen in quantum dots [13-17], including (i) a narrow conductance peak at zero source-drain bias that forms at low temperature, (ii) collapse of conductance data onto a single function—an empirical Kondo-like form—over a range of gate voltages in the vicinity of the 0.7 feature using a single scaling parameter (which we designate the Kondo temperature), (iii) correspondence between the Kondo scaling factor and the width of the zero-bias peak, and (iv) splitting of the zero-bias peak in a magnetic field. An important difference between the Kondo effect in quantum dots and the present situation is that the QPC has no obvious localized state. A speculation about the possibility of a Kondo state in a QPC was made recently [18], but to our knowledge no concrete theory of such an effect has yet been formulated.

Measurements were made on five QPCs of slightly different lengths and widths fabricated on a high-quality delta-doped GaAs/AlGaAs heterostructure with a two-dimensional electron gas (2DEG) 100 nm below the surface. Data is from the device pictured in Fig. 1(c), though all devices displayed qualitatively similar behavior. Measurements were carried out in a dilution fridge with an estimated base electron temperature of ~ 80 mK. The differential conductance, $g = dI/dV$, was measured as a function of gate voltage, $V_g$, temperature, T, in-plane magnetic field, B, and dc source-drain bias, $V_{sd}$, using standard ac lockin techniques with a small ac bias voltage, $|V| < 10\mu V$ [19].

The linear-response conductance (i.e., g around $V_{sd}$ ~ 0) exhibits a characteristic evolution from spin-degenerate plateaus at B = 0, appearing at integer multiples of $2e^2/h$, into spin-resolved plateaus at integer multiples of $e^2/h$ in high field [Fig. 1(b)]. A remnant of the spin-resolved plateau remains at B = 0 and T = 80 mK as a barely-visible shoulder below the $2e^2/h$ plateau. As the temperature is increased, conductance at this shoulder decreases and a clear plateau near $0.7(2e^2/h)$ forms [Fig. 1(a)]. Note that while the 0.7 structure becomes stronger at elevated temperatures, the spin-degenerate plateaus at multiples of $2e^2/h$ become more washed out. Stated another way, as the temperature is *lowered*, the spin degenerate plateaus at $2e^2/h$ sharpen up, while the plateau at $0.7(2e^2/h)$ rises to the "unitary limit" value of $2e^2/h$, and thus disappears.

Nonlinear transport data shown in Figs. 1(d-f) as a series of traces of conductance g versus $V_{sd}$ measured at different gate voltages (not offset) further emphasize the similarity between the spin-resolved plateaus at B = 8 T and the extra plateau in the first mode at B = 0. In this representation, plateaus in g(V$_g$) appear as an accumulation of traces, seen for instance in the (well-understood) "half-plateaus" [20,21] at higher bias (V$_{sd}$ > ~ 0.5 mV) at g ~ 1/2, 3/2 and 5/2 (in units of $2e^2/h$), labeled in Fig. 1(d). The quantized linear-response plateaus are visible as accumulated traces around zero bias at multiples of $2e^2/h$ and, in the B = 8 T data [Fig. 1(f)], also at odd multiples of $e^2/h$. Note the distinctive wing shape of the spin-resolved plateaus, rising with increased bias from ~ 0.5 to ~ 0.8 of the distance between the spin-degenerate plateaus, with a transition around |V$_{sd}$| ~ 0.2 mV. This width is consistent with the Zeeman splitting at 8 T ($g^*\mu_B B$ ~ 25 µV/T using $|g^*| = 0.44$).

Within the first subband (g < $2e^2/h$) the nonlinear data for B = 0 [lower region of Fig. 1(e)] look strikingly similar to the B = 8 T data [lower region of Fig. 1(f)], including the wing shape of the extra plateau that extends out from the 0.7 feature. Note that higher subbands at B = 0 [upper region of Fig. 1(e)] do not show extra plateaus. The overall impression given by comparing Figs. 1(e) and 1(f) is consistent with the linear transport data: The extra plateau that starts at ~ 0.7($2e^2/h$) and extends to ~ 0.8($2e^2/h$) at high-bias appears to result from a splitting of spin bands, leading to transport signatures in the lowest mode that greatly resemble the situation at 8 T, where spin degeneracy is explicitly lifted in all modes by the applied field.

The low temperature nonlinear data [Fig. 1(d)] show a significant additional feature compared to the higher temperature data [Fig. 1(e)]: a narrow peak in conductance around V$_{sd}$ = 0 for the whole range 0 < g < $2e^2/h$. This zero-bias anomaly (ZBA) forms as the temperature is lowered, as seen in Fig. 2(a). The ZBA is closely linked to the disappearance of the 0.7 structure at low temperature: comparing Figs.1(d) and 1(e), one sees that it is precisely this ZBA peak that lifts the 0.7 plateau toward $2e^2/h$.

The formation of a zero-bias conductance peak, and the associated enhancement of the linear conductance up to the unitary limit ($2e^2/h$) at low temperature is reminiscent of the Kondo effect seen in quantum dots containing an odd number of electrons [13-17,22,23]. Guided by this similarity, we consider a scaling of the temperature dependence of the conductance using a single scaling parameter which we designate the Kondo temperature, $T_K$. Experimentally, we find that this single parameter allows data from a broad range of gate voltages [Fig. 2(b), inset] to be scaled onto a single curve as a function of scaled temperature $T/T_K$ [Fig. 2(b)]. Moreover this scaled curve appears well described by a modified expression for the Kondo conductance,

$$g = e^2/h \, [f(T/T_K) + 1] \qquad (1)$$

where $f(T/T_K)$ is a universal function for the Kondo conductance (normalized to $f(0) = 1$) [24] well approximated by

$$f(T/T_K) \sim [1 + (2^{1/S} - 1)(T/T_K)^2]^{-S} \qquad (2)$$

with $s = 0.22$ [15] . Equation (1) differs from the form that has been used for quantum dots [15], $g = (2e^2/h)f(T/T_K)$, by the addition of a constant $e^2/h$ term that sets the high-temperature limit to $e^2/h$ rather than zero, and by fixing the prefactor of $f(T/T_K)$ to one. The motivation for adding a constant term of $e^2/h$ to the usual dot form is primarily empirical: allowing a prefactor and added constant to be fit parameters along with $T_K$ consistently gave values close to these; locking both values to one had essentially no effect on the fit values for $T_K$ [25]. Additionally, the fact that a QPC does not show Coulomb blockade motivates one to consider functional forms that do not go to zero for $T \gg T_K$.

The one-parameter fit to the g(T) data using Eqs. (1) and (2) yields values for $T_K$ that increase exponentially with the gate voltage, $\ln(T_K) \sim a(V_g - V_g^o)$, with a = 0.18 (for $T_K$ in Kelvin and $V_g$ in mV) obtained from a best fit line to $\ln(T_K)$ [Fig. 2(c)]. The exponential dependence of $T_K$ on $V_g$ is perhaps not surprising given that for quantum dots [26] $T_K \sim \exp[\pi\varepsilon_0(\varepsilon_0+U)/(\Gamma U)]$ (neglecting nonexponential prefactors) depends exponentially on $\varepsilon_0$, the energy of the bound spin relative to the Fermi energy of the leads, $(\varepsilon_0+U)$, the energy to the next available state, and $\Gamma$, the energy broadening due to coupling to the reservoirs.

A characteristic feature of the low-temperature Kondo regime ($T < T_K$) in quantum dots is that the ZBA peak is split by $2g^*\mu_B B$, twice the Zeeman energy, upon application of an in-plane magnetic field as long as $g^*\mu_B B > \sim T_K$ [14,17,23]. In the QPC, we find the ZBA peak does not split uniformly over the full range $0 < g < 2e^2/h$, as seen in Fig. 2(d). At conductances in the vicinity of g ~ 0.7 clear splitting is seen, consistent with $2g^*\mu_B B$ (i.e., the splitting is roughly linear in field up to ~ 3 T, and consistent with a g-factor that is ~1.5 times the bulk GaAs g-factor). At higher conductances the ZBA peak does not split with B, but merely collapses [top curves in Fig. 2(d)]. The absence of splitting at higher conductances is expected, since $2g^*\mu_B B < T_K$ in this regime. The less prominent splitting of the ZBA at low conductance [bottom traces in Fig. 2(d)] may result from a lower Kondo temperature, such that $T \sim T_K$, at these gate voltages.

Another well known feature of the Kondo effect as seen in quantum dots is that the width of the ZBA is set by $T_K$ rather than the larger level-broadening scale $\Gamma$ [16]. As seen in Fig. 3(a), the width of the ZBA [Fig. 3(a), inset] is roughly constant for $g < 0.7$, ($V_g < \sim 490$ mV). At $g \sim 0.7$, the ZBA first narrows significantly, by ~30%, then broadens as g approaches $2e^2/h$. The ZBA peak width is very close to $2kT_K/e$ [squares in Fig. 3(a)] for $g > 0.7$, where values of $T_K$ can be extracted. Relations between the ZBA width and the values of $kT_K/e$ with a similar prefactor of $\sim 2$ is observed in quantum dots [16] and nanotubes [17].

A related correspondence between $T_K$ and the applied bias voltage is seen in Fig. 3(b). The colorscale reflects the derivative of g with respect to gate voltage ($dg/dV_g$) and so emphasizes the transitions between plateaus (seen as red/yellow bands), while the plateaus appear black. The 0.7 structure appears as an extra pair of transitions, symmetric in $V_{sd}$, with a distinctive downward curvature as they approach the "origin" at $V_{sd} = 0$ and $V_g \sim -500$ mV. Crossing these features into the central black diamond marks the transition from the extra plateaus at $\sim 0.8$ $(2e^2/h)$ [seen in Fig. 1(d-f)] to the $2e^2/h$ plateau. These extra transitions are greatly diminished or absent in higher subbands [5,7]. Superimposed on the color plot in Fig. 3(b) are the Kondo temperatures taken from Fig. 2(c), plotted at an equivalent "Kondo bias voltage" $V^K_{sd} = kT_K/e$ and at the $V_g$ where that $T_K$ was measured. The alignment of these points with the extra transitions suggests that an applied bias exceeding $V^K_{sd}$ destroys the correlated Kondo-like state causing the conductance to drop to the high-bias value of the extra plateau, ~ $0.8(2e^2/h)$.

The Kondo-like correlation picture sheds some light on several puzzling aspects of the 0.7 structure itself, namely, the nature of the spin splitting, why the plateau is typically above g = 0.5, and the fact that the plateau disappears at low temperature. In order for a Kondo-like effect to appear in a QPC, the splitting between spin bands must be *dynamic* rather than a frozen spin polarization at zero field. The explanation for g > $0.5(2e^2/h)$ on the plateau is suggested by the scaling plot [Fig. 2(b)]: the proposed high-temperature limit is g = $0.5(2e^2/h)$. However, the Kondo contribution to g decays very slowly (logarithmically) with increasing temperature, leaving a significant residual enhancement. The recovery of the conductance to the unitary limit of $2e^2/h$ with decreasing temperature is a natural feature of a fully developed Kondo effect.

We thank H. Bruus, L. Glazman, B. Halperin, L. Levitov, and Y. Meir for useful discussions. Supported by ARO-MURI DAAD-19-99-1-0215, the Lucent Technologies Foundation GRPW Program (H.J.L.), and the Harvard Society of Fellows (D.G.-G.).

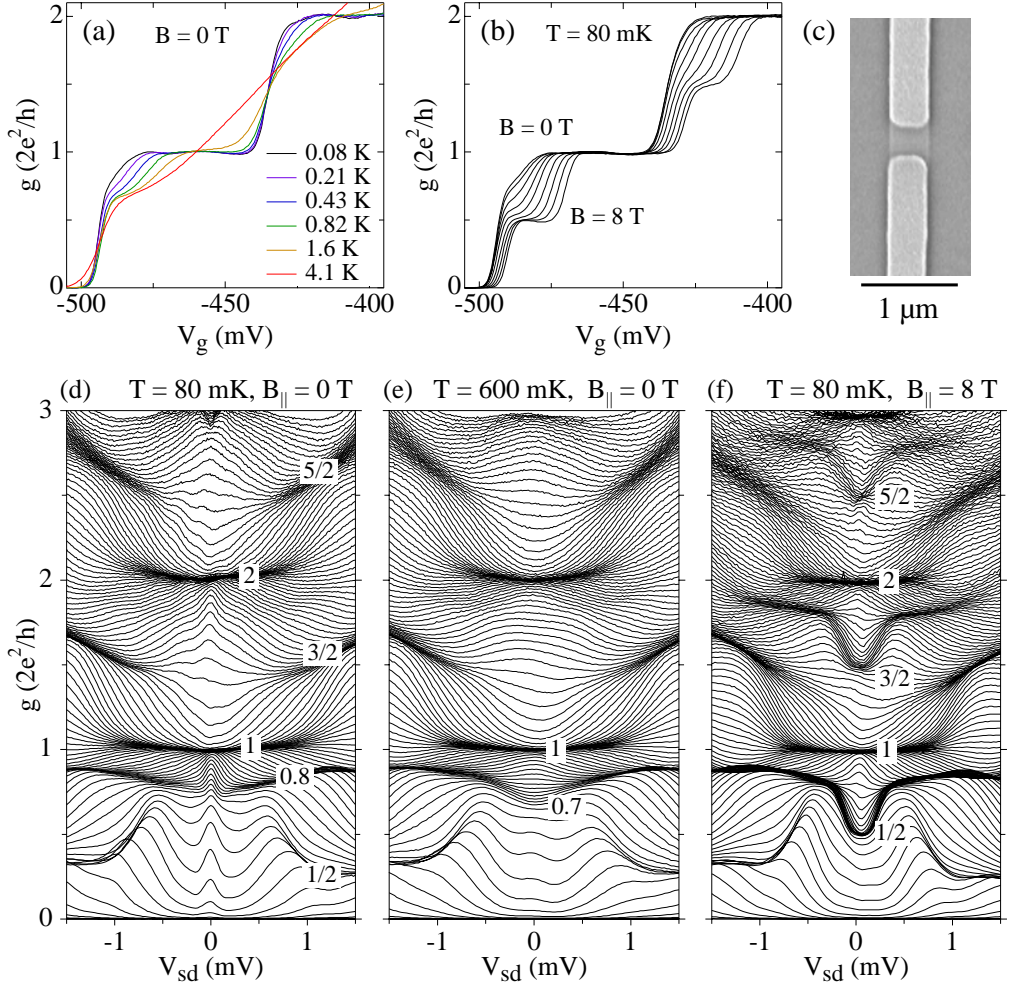

**FIG. 1.** (a) Linear conductance (g = dI/dV, around $V_{sd} \sim 0$) versus gate voltage, $V_g$, at B = 0 for several temperatures. The extra plateau at $\sim 0.7(2e^2/h)$ appears with increasing temperature while the plateaus at multiples of $2e^2/h$ become less visible due to thermal smearing. (b) Linear response g versus $V_g$, for in-plane field B from 0 to 8 T in 1 T steps, showing spin-resolved plateaus at odd multiples of $e^2/h$ at high fields. (c) Micrograph of the device used. (d-f) Nonlinear differential conductance g = dI/dV as a function of dc source-drain bias voltage, $V_{sd}$, with each trace taken at a fixed gate voltage. Plateaus in $g(V_g)$ appear as dense regions where many traces accumulate (traces are not slid). (d) Nonlinear g at 80 mK, B = 0, at $V_g$ intervals of 1.25 mV. Plateaus at multiples of $2e^2/h$ around $V_{sd} \sim 0$ and half-plateaus at odd multiples of $e^2/h$ at high bias [20] are visible. A zero-bias anomaly (ZBA) is present only at low magnetic field and low temperatures. At high bias, an extra plateau appears at $g \sim 0.8(2e^2/h)$ (e) Nonlinear g at 600 mK, B = 0, at $V_g$ intervals of 1.0 mV. Note absence of a ZBA and an accumulation of lines at $g \sim 0.7(2e^2/h)$ around $V_{sd} \sim 0$ that merges with the high-bias plateau at $0.8(2e^2/h)$. (f) Nonlinear g at 80 mK, B = 8 T, at $V_g$ intervals of 1.2 mV. Spin-resolved plateaus at odd multiples of $e^2/h$ around $V_{sd} \sim 0$ merge with high-bias plateaus at $0.8(2e^2/h)$, $1.8(2e^2/h)$, and $2.8(2e^2/h)$. The high-bias feature at $0.8(2e^2/h)$ looks similar to that in the B = 0 data.

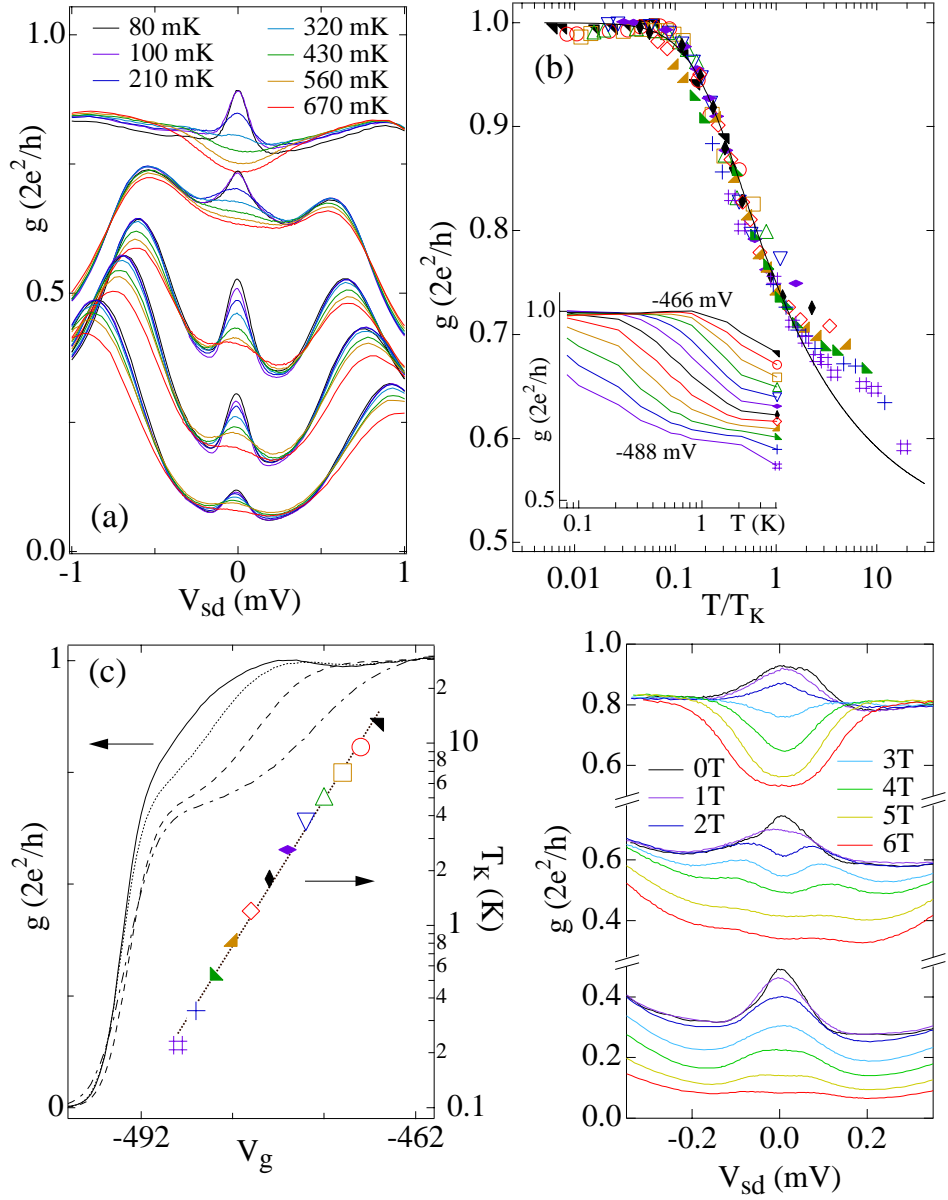

**FIG. 2. (a)** Temperature dependence of the zero-bias anomaly (ZBA) for different gate voltages, at temperatures from 80 mK to 670 mK. **(b)** Linear conductance, g, as a function of scaled temperature $T/T_K$ where $T_K$ is the single fit parameter in Eqs. 1, 2. Symbols correspond to gate voltages shown in inset. **Inset:** Linear conductance as a function of unscaled temperature, T, at several $V_g$. **(c)** $T_K$ (right axis) obtained from the fits of $g(T/T_K, V_g)$ to Eqs. 1, 2, along with the conductance (left axis) at temperatures of 80 mK (solid line), 210 mK (dotted), 560 mK (dashed), and 1.6 K (dot-dashed). **(d)** Evolution of the ZBA with in-plane B, at $V_g$ corresponding to high, intermediate, and low conductance. Splitting is clearly seen in the intermediate conductance data (see text). Data in (d) were measured with zero perpendicular field.

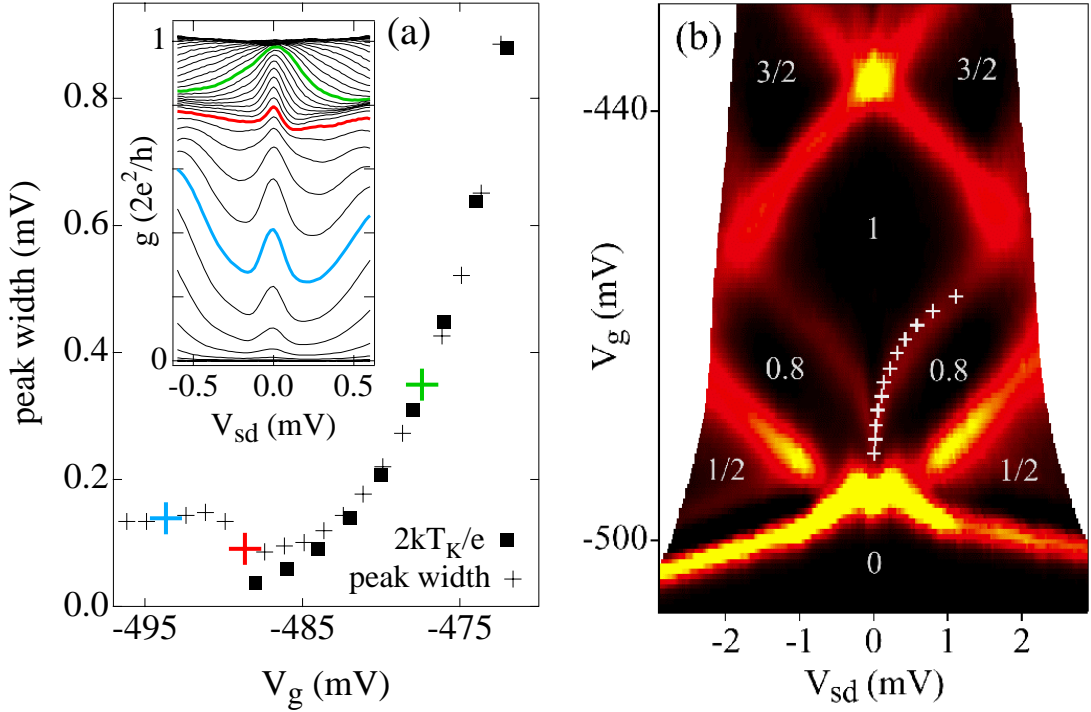

**FIG. 3 (a)** Widths of the ZBA peak (crosses), defined as the full-width at half-max from the local minimum on the left side, for gate voltages corresponding to $g < 2e^2$h. The values $2kT_K/e$ (squares) are shown for the range of $V_g$ where $T_K$ could be extracted, $0.7(2e^2/h) < g < 2e^2/h$. Colored crosses correspond to colored traces in the inset. **Inset**: Nonlinear g from Fig. 1(d) $V_{sd} \sim 0$. **(b)** Numerical derivative of conductance (colorscale) with respect to gate voltage, $dg/dV_g$ as a function of $V_{sd}$ and $V_g$ at B = 0 and T = 80 mK. Black regions ($dg/dV_g \sim 0$) correspond to plateaus in $g(V_g)$, Colors mark transitions between plateaus in $g(V_g)$. A simple spin-degenerate model of a QPC would show a vertical sequence of X's. The black diamond-shaped regions, formed between subsequent X's, correspond to conductance plateaus at multiples of $2e^2/h$. The experimental data show the plateau at $1(2e^2/h)$, high-bias half-plateaus at 1/2, 3/2, and extra plateaus associated with the 0.7 structure that rise to $0.8(2e^2/h)$ at high bias. Kondo voltage, $kT_K/e$ (white crosses), are superimposed on the color scale plot at several values of $V_g$ with no adjustment. The Kondo voltage agrees well with the position of the transition from the $2e^2/h$ central diamond to the extra plateaus at $0.8(2e^2/h)$, including the noticeable curvature near $V_{sd} = 0$ at $V_g \sim -500$ mV.